\documentclass[11pt,a4paper]{article}
\usepackage{xcolor}
\usepackage{graphics}
\usepackage{graphbox}
\usepackage{authblk}
\usepackage{multicol}
\usepackage[margin=0.6in]{geometry}
\usepackage{caption}
\usepackage{braket}
\usepackage{amsfonts,amsmath,amssymb}
\usepackage{url}
\begin{document}
	\title{Quantum repeaters in space}
	\date{}
	\author{Carlo Liorni$^{1*}$, Hermann Kampermann$^1$, Dagmar Bru\ss$^1$}
	\affil{$^1$Heinrich-Heine-Universit\"at, Institut f\"ur Theoretische Physik III, Universit\"atsstr. 1, 40225, D\"usseldorf, Germany. Correspondence and requests for materials should be addressed to C.L. (email: liorni@uni-duesseldorf.de)}

	\maketitle
	
	\begin{abstract}
		
		Long-distance entanglement is a very precious resource, but its distribution is very difficult due to the exponential losses of light in optical fibres.
		A possible solution consists in the use of quantum repeaters, based on entanglement swapping or quantum error correction. 
		Alternatively, satellite-based free-space optical links can be exploited, achieving better loss-distance scaling. 
		We propose to combine these two ingredients, quantum repeaters and satellite-based links, into a scheme that allows to achieve entanglement distribution over global distances with a small number of intermediate untrusted nodes. The entanglement sources, placed on satellites, send quantum states encoded in photons towards orbiting quantum repeater stations, where entanglement swapping is performed. 
		The performance of this repeater chain is assessed in terms of the secret key rate achievable by the BB-84 cryptographic protocol. We perform a comparison with other repeater chain architectures and show that our scheme is superior in almost every situation, achieving higher key rates, reliability and flexibility. Finally, we analyse the feasibility of the implementation in the mid-term future and discuss exemplary orbital configurations.
		The integration of satellite-based links with ground repeater networks can be envisaged to represent the backbone of the future Quantum Internet.  
		
	\end{abstract}

	\newpage
	
	\begin{multicols}{2}
	
	Entanglement distribution between very distant parties allows several interesting quantum-enabled protocols to be performed, in the fields of quantum communication \cite{gisinreview,krenn}, metrology \cite{metrology1,metrology2,metrology3} and distributed computation \cite{distr1,distr2}.  However, achieving this task over global distances (thousands of km) is very daunting. The standard carrier of quantum information is light, sent through optical fibres. The exponential losses experienced during the propagation limit the achievable distances to $\sim200$ km in practice. The concept of a Quantum Repeater (QR) \cite{repeater1,rep5,rep1,rep2,rep3,rep4} has been introduced to counter this problem. Such a device allows, using Quantum Memories (QMs) \cite{qmem1} and protocols based on Entanglement Swapping (ES) or quantum error correction \cite{ecrep}, to connect several elementary links and enlarge the achievable distance.
	
	An alternative solution is represented by satellite-relayed free-space channels. Satellite-to-ground optical links for quantum communication have already been proven to be feasible with current technology \cite{Liao,pan2,pan3,pan4,bonato,jennewein1,liorni}. They allow, in the double down-link configuration, to share entanglement between two ground stations, at distances that far exceed what can be achieved with direct fibre transmission. Low Earth Orbit (LEO, altitude $\lesssim$ 2000 km) satellites are preferred, because of the lower cost and the shorter distance between the satellite and the ground stations, which reduces the overall loss in the channel. However, the maximum distance between the ground stations is limited to $\{1500-2000\}$ km, due to the additional losses encountered at low elevation angles. This aspect makes intercontinental quantum communication not feasible with such a scheme. 
	
	\noindent
	\begin{minipage}{\linewidth}
		\centering
		\includegraphics[width=0.9\columnwidth]{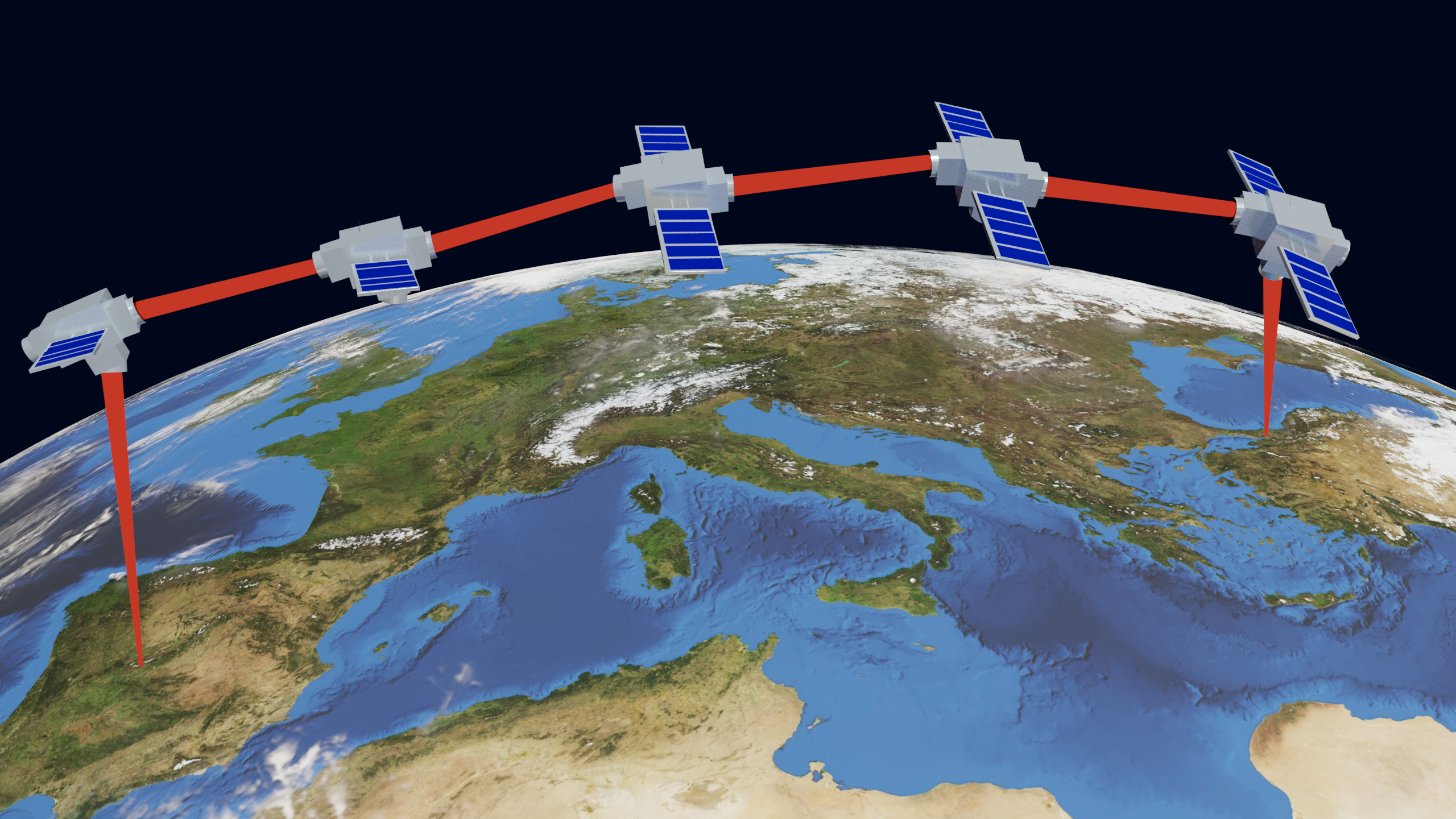}
		\captionof{figure}{Pictorial representation of the scheme proposed in this paper for long-distance entanglement distribution, based on orbiting quantum repeater stations.}
		\label{fig:orbiting}
	\end{minipage}
	
	Through quantum repeaters, few of these satellite links can be chained together to reach global distances. In this work we propose and study the scheme pictured in Fig.~\ref{fig:orbiting}, in which entanglement sources and quantum repeaters are placed on board of satellites, orbiting around the Earth in the \emph{string of pearls} configuration. This allows to connect two users on the ground via free-space optical links outside the atmosphere, achieving far superior distance-to-loss ratio with respect to the standard fibre-based implementation. In this way, a small number of intermediate nodes is enough to achieve entanglement distribution over global distances at a reasonable rate.
	
	We focus in the following on a specific application of entanglement distribution, namely Quantum Key Distribution (QKD). The secret key rate turns out to be a good measure of the effectiveness of the quantum repeater link \cite{abruzzo}. 
	We compare the performance of the newly-proposed scheme with two other quantum repeater configurations based on entanglement swapping. The nomenclature used in the remainder of the paper is the following, also schematically represented in Fig.~\ref{fig:3schemes}: scheme OO (Orbiting sources Orbiting repeaters) is our proposal, scheme GG (Ground sources Ground repeaters) is the fibre-based one and scheme OG (Orbiting sources Ground repeaters) is the solution proposed in \cite{jennewein2}, where the quantum repeater stations are on the ground. We show that our configuration ensures, at the expense of additional technical difficulty, better performance, reliability and availability, as discussed in the next section.

	\noindent 
	\begin{minipage}{\linewidth}
		\centering
		\includegraphics[width=0.99\columnwidth]{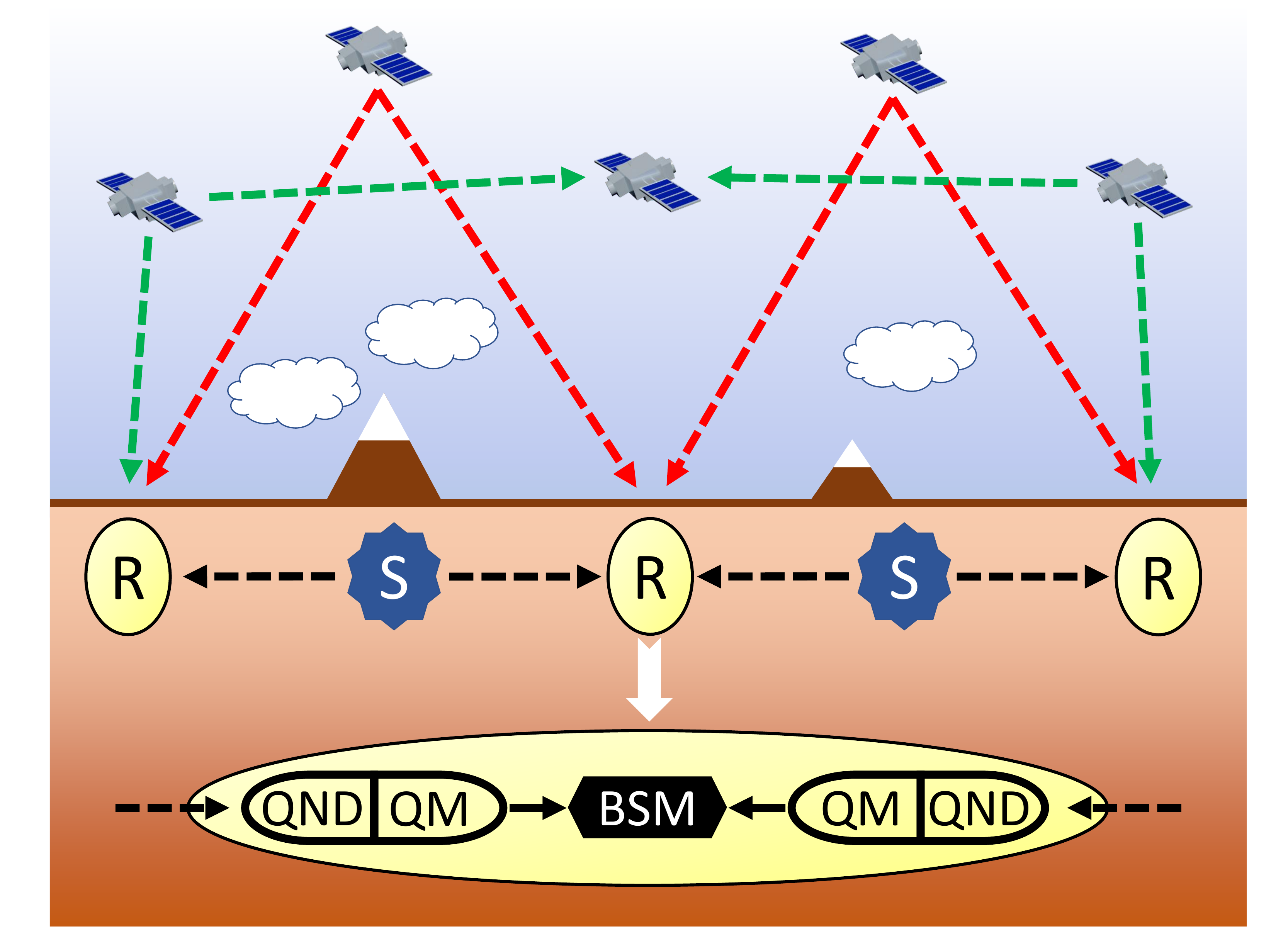}
		\captionof{figure}{Schematic comparison between the satellite-based scheme OO (green arrows), the standard fibre-based implementation (GG, in black) and the scheme studied in \cite{jennewein2} (OG, in red). Here S represent entanglement sources and R quantum repeater stations. The incoming photons are heralded by Quantum Non-Demolition meaurement devices (QND) and the quantum information is loaded into Quantum Memories (QM). Finally, the quantum states are read and a Bell State Measurement (BSM) is performed, as part of the entanglement swapping protocol.}
		\label{fig:3schemes}
	\end{minipage}
	
	In Sec.~\ref{results1} we quantitatively estimate the performance of the different schemes in terms of achievable secret key rate and compare them. Afterwards, in Sec.~\ref{results2}, we discuss pros and cons of the proposed satellite-based scheme and then analyse two exemplary orbital configurations in Sec.~\ref{examples}. The results are briefly summarized and discussed in the conclusion, Sec.~\ref{discussion}. Additional details on the simulations can be found in Sec.~\ref{methods}, regarding the error model and the contribution of environmental photons, the analysis of the orbits, the estimation of the satellite link transmittance and the values of the parameters used.

	\section{Results}
	\label{results}
	
	\subsection{Secret key rate and comparison}
	\label{results1}
	
	The quantum repeater architecture is designed as follows \cite{rep0}. The total link of length L between the two communicating parties A and B is divided into $2^n$ elementary links of length $l_0=L/2^n$. Quantum repeaters are placed at the connections between adjacent elementary links, while entanglement sources are in their central points (Fig.~\ref{fig:3schemes}). The latter produce bipartite entangled states (in the following we will consider qubit pairs), encoded in some degree of freedom of a pair of photons, that are then injected in the adjacent elementary links. The quantum repeaters consist of 3 main devices. First of all Quantum Non-Demolition (QND) measurement devices herald the arrival of a photon from the elementary link. The quantum state encoded in the heralded photons is then loaded and stored in quantum memories. When both memories are full, a joint Bell State Measurement (BSM) is performed and the result broadcast. This entanglement swapping procedure allows to connect two adjacent entangled pairs and, repeated in a recursive and hierarchical way, to gradually extend the entanglement (see \cite{abruzzo} for details). In consecutive \emph{nesting levels}, the distance between the subsystems composing the entangled pairs will be doubled, with $n$ the maximal nesting level. After $n$ successful steps the entanglement is shared between the end points of the chain, the parties A and B.
	
	In the case of scheme OO, the elementary links consist of double inter-satellite links and hybrid inter-satellite/down-link at the end points. In scheme GG, instead, they consist of optical fibres, whose transmittance \mbox{$\eta_f(l)=10^{-\alpha l/10}$} decreases exponentially with the length $l$, where the attenuation parameter is $\alpha=0.17$ dB/km at 1550nm. Scheme OG on the other hand comprises double down-links from the satellites towards two adjacent receiving stations on the ground (as in Fig.~\ref{fig:3schemes}).
	We discuss the losses introduced by such satellite links in Sec.~\ref{methods2}. 
	After an entangled pair is successfully shared between the parties A and B, it can be used for any quantum information protocol, in particular QKD. In this cryptographic primitive the two parties are connected by an insecure quantum link, the repeater chain, and by an authenticated classical channel. An eavesdropper can tamper on the classical channel and freely interact with the states sent over the quantum channel. The parties have to devise a protocol that either creates a private key or aborts. A generic protocol usually comprises the exchange of quantum states with successive measurements in random bases, base sifting, parameter estimation, error correction and privacy amplification. In the following we apply the well known asymmetric BB84 protocol \cite{bb84}, in which the quantum states are prepared and measured in the bases defined by the eigenstates of the $X$ and $Z$ Pauli operators on qubits. For the security analysis \cite{abruzzo} we assume that the whole quantum repeater chain is untrusted, so, not only the quantum channels, but also the sources on the satellites and the repeater stations can be in the eavesdropper's hands.
	We do not focus on any specific implementation regarding the encoding of the quantum information in the single photons. For the satellite-based schemes polarization is feasible \cite{Liao,pan2,pan3,pan4} and promising, so we base the error model on this assumption. In any case, the choice of the encoding is strongly linked with the choice of the quantum memory architecture and material.
	The secret key rate depends on both the repeater rate and the quality of the final shared entangled state.
	It is estimated in the limit of an infinitely long key, based on the considerations in \cite{abruzzo}, by:
	
	\begin{equation}
	\label{eq:key}
	R_{QKD}^{BB84}=R_{\text{rep}} \ P_{\text{click}} \ R_{\text{sift}} \ r_{\infty}^{BB84} \ .
	\end{equation}
	
	In the expression above, $R_{\text{rep}}$ represents the entanglement distribution rate of the repeater chain, $P_{\text{click}}$ the double detection probability, $R_{\text{sift}}$ the sifting ratio (assumed equal to 1 in our asymmetric and asymptotic protocol) and $r^{BB84}_{\infty}$ the BB-84 secret fraction: 
	
	\begin{equation}
	\label{eq:rep}
	R_{\text{rep}}=\frac{1}{T_0} P_0 \ P_{QND}^2 \ P_W^2 \bigg( \frac{2}{3} P_{ES} \ P_R^2 \bigg)^n
	\end{equation}
	
	\begin{equation}
	\label{eq:P}		
	P_{\text{click}}=\eta_d^2 \qquad r_{\infty}^{BB84}=1-h(e_Z)-h(e_X) \ .
	\end{equation}
	
	In Eq.~(\ref{eq:rep}), the quantity $1/T_0$ represents the intrinsic repetition rate of the repeater architecture. We assume here that the memories used here are multi-mode \cite{deRiedmatten,multimode} (see \cite{jennewein2} for additional discussions) so that we can avoid to wait acknowledgement from the adjacent stations that the photons have been received, before proceeding with the protocol or emptying the memory. This allows us to fix $T_0=1/R_s$, with $R_s$ the repetition rate of the source. The memory bandwidth of the chosen material limits the maximum repetition rate, that we fix to 20 MHz for the following simulations \cite{jennewein2,lukinmemo}. $P_0$ is the transmittance of the elementary links which depends on the scheme under study. We identify with $P_0$ the average of the link transmittance over one fly-by of the satellite for schemes OO and OG. $P_{QND}$, $P_W$ and $P_R$ are, respectively, the efficiency of the QND measurement and the writing and reading efficiency of the quantum memory. $P_{ES}$ is the success probability of the single entanglement swapping process (we refer to Sec.~\ref{methods1} and \cite{abruzzo} for details). 
	The term $2/3$ is connected with the average amount of time that one has to wait until entangled pairs in adjacent segments of the repeater chain are successfully generated. It arises due to a commonly employed approximation valid for small $P_0$, which is always valid in the cases under study (we refer to \cite{vanlock} for further details and the exact solution). In Eq.~(\ref{eq:P}), $\eta_d$ is the efficiency of the detectors used for the final measurement of the photons. The secret fraction $r_{\infty}^{BB84}$ depends, through the binary entropy $h(p)=-p \ \text{log}_2(p)-(1-p)\text{log}_2(1-p)$, upon the error rates in the $X$ and $Z$ bases, $e_X$ and $e_Z$. In our simulations they are estimated tracking the evolution of the state of the entangled pairs throughout the ES process, starting from noisy elementary pairs. In a practical experiment these error rates are the result of the parameter estimation stage, in which the parties make public a small subset of their measurement results and compare them.
	
	In the following we assume two-qubit systems and we consider, without loss of generality, an entangled state $\rho_{AB}$ diagonal in the Bell basis
		\begin{eqnarray}
		\rho_{AB}=p_{\phi^+}\ket{\phi^+}\bra{\phi^+}+p_{\phi^-}\ket{\phi^-}\bra{\phi^-}& \nonumber \\  
		+p_{\psi^+}\ket{\psi^+}\bra{\psi^+}+p_{\psi^-} \ket{\psi^-}\bra{\psi^-}&
		\end{eqnarray}
	with $p_{\phi^+}+p_{\phi^-}+p_{\psi^+}+p_{\psi^-}=1$ and the Bell states $\ket{\phi\pm}=(\ket{11}\pm \ket{00})/\sqrt{2}$ and $\ket{\psi^\pm}=(\ket{10}\pm \ket{01})/\sqrt{2}$. It is possible to apply appropriate local twirling operations that transform an arbitrary two-qubit quantum state in a Bell diagonal state, without compromising the security of the protocol \cite{renner1}.
	This structure of the state simplifies the analysis because it can be shown that starting from two Bell-diagonal pairs, the resulting state after entanglement swapping between two sub-systems is still Bell diagonal and the new coefficients $p_{\phi^+}',p_{\phi^-}',p_{\psi^+}',p_{\psi^-}'$ can be readily computed \cite{abruzzo}. Then, the error rates along the $X$ and $Z$ directions can be simply written as 
	\begin{equation}
	e_X=p_{\phi^-}+p_{\psi^-} \quad e_Z=p_{\psi^+}+p_{\psi^-} \ .
	\end{equation}

	The Bell-diagonal state received by the adjacent repeater stations is assumed to be, without loss of generality, a depolarized state of fidelity $F$ with respect to $\ket{\phi^+}$
	\begin{eqnarray}
	\label{rho0}
	&&\rho=\rho^{\text{dep}}(F)=F \ket{\phi^+}\bra{\phi^+} \\
	&&+ \frac{1-F}{3}(\ket{\psi^+}\bra{\psi^+}+\ket{\psi^-}\bra{\psi^-}+\ket{\phi^-}\bra{\phi^-}) . \quad \nonumber
	\end{eqnarray}
	The fidelity $F$ accounts for the initial fidelity of the entanglement sources on the satellites and for the noise model that describes the channel. A depolarized state is a natural choice as it corresponds to a common and generic noise that well suits the problem under study and, moreover, any two-qubit mixed quantum state can be reduced to this form using some (previously mentioned) local twirling operations \cite{bennett1}.
	
	In the presence of environmental photons entering the receiver, the probability that the detection was due to a signal photon from the adjacent satellite can be estimated as	
	\begin{equation}
	P_{s}=\frac{N_{s}}{N_{s} + N_{n}} \ ,
	\end{equation}
	where $N_{s}$ represents the number of signal photons per time window that we expect to observe (proportional to the transmittance of the channel) and $N_{n}$ is the expected number of environmental photons in the same time window. Now, with the assumption that environmental photons are unpolarized and uncorrelated to the signal photons, the final state the repeater stations receive is modelled as a mixture of the initial state sent by the sources $\rho_0$ with the completely mixed state
	\begin{equation}
	\rho= P_{s1}P_{s2} \rho_0 + (1-P_{s1}P_{s2}) \frac{\mathbb{I}}{4} \ ,
	\end{equation}
	where $\mathbb{I}$ is the $4\times4$ identity matrix and $P_{s1}$ and $P_{s2}$ refer to the receiving telescopes of the adjacent repeater stations.
	
	Introducing the definition of the initial state \mbox{$\rho_0=\rho^{\text{dep}}(F_0)$} with the initial fidelity $F_0$ and writing the completely mixed state in the Bell basis we obtain, after comparison with Eq.~(\ref{rho0}),
	\begin{equation}
	\label{FFF}
	F= P_{s1} P_{s2} F_0 + (1-P_{s1} P_{s2}) \frac{1}{4} \ .
	\end{equation}
	In Sec.~\ref{methods1} we show how to estimate the probabilities $P_{s1}$ and $P_{s2}$ in the different cases and which are the most important sources of environmental photons. The fibre-based implementation is substantially immune to this problem and we neglected further sources of error like basis misalignment, so the state that the repeater stations received is actually $\rho_0$.
	
	We point out that no entanglement distillation \cite{distillation} is performed in the protocol analysed here. If high quality gates for the implementation of entanglement distillation are available, this operation may allow to get higher key rates and reduce the threshold on the initial fidelity of the pairs and the noise filtering. 
	
	Now we discuss the results of the comparison between scheme OO and the other configurations. The parameters employed for the simulations are given in Tab.~\ref{paraTab} of the Methods section. In particular, for schemes OO and OG, we assume the radii of the main optical elements to be 25 cm for the emitters and 50 cm for the receivers, values compatible with mini-satellites and standard optical ground stations. The transmittance of the free-space links is estimated assuming an imperfect Gaussian beam and a simple model for the atmospheric extinction (more details in Sec.~\ref{methods2}). Regarding detector and quantum memory efficiencies, we assumed rather conservative values, that either have already been achieved separately in different implementations or are expected to be reached in the near future \cite{jennewein2}.
	
	In Fig.~\ref{fig:long123} we show the secret key rate, see Eq.~(\ref{eq:key}), as a function of the total distance between the parties for several interesting configurations of schemes OO, GG and OG, in the range $[1000,20000]$ km.

	\noindent
	\begin{minipage}{\linewidth}
		\centering
		\includegraphics[width=1\columnwidth]{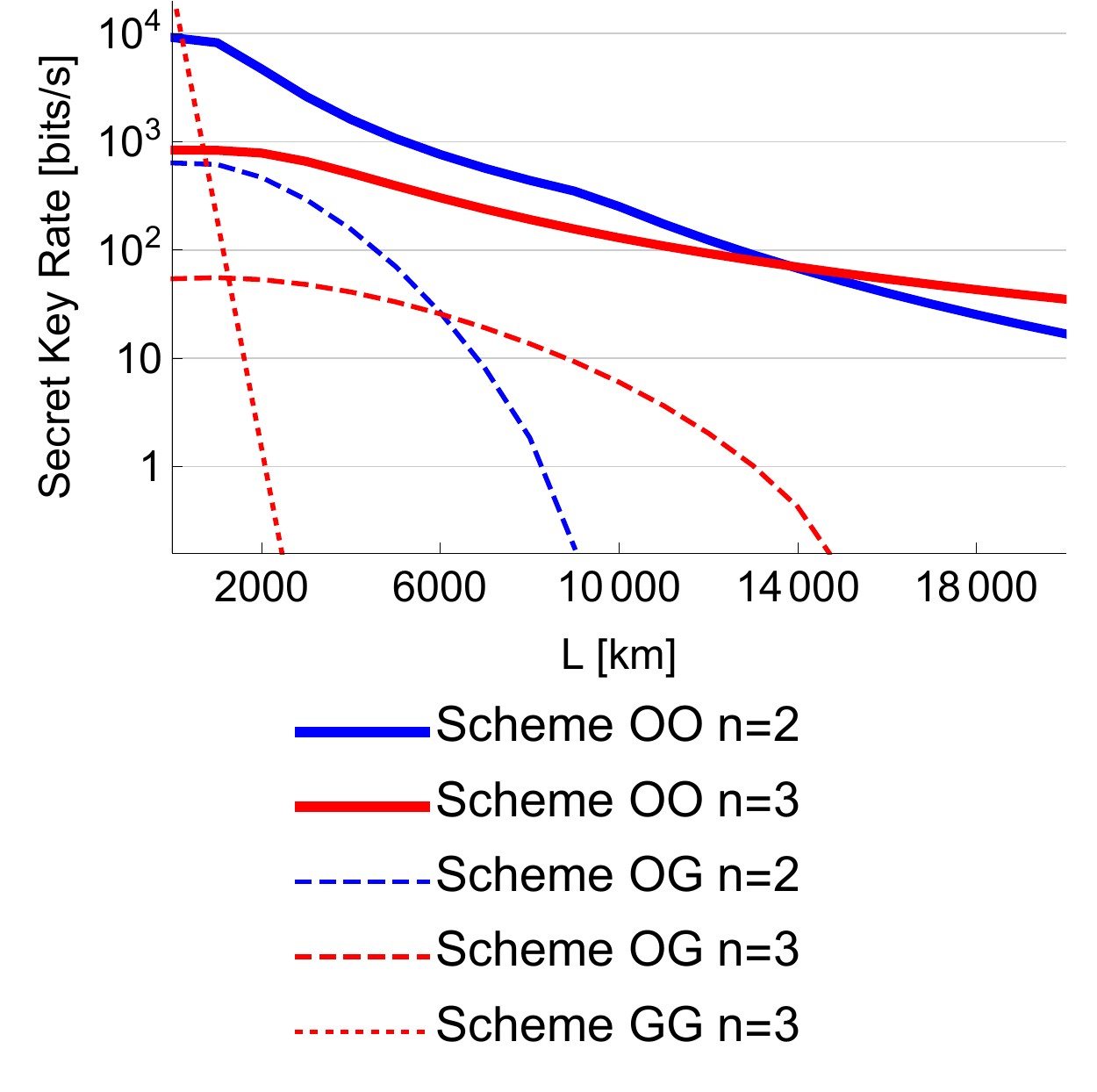}
		\captionof{figure}{Secret key rate, see Eq.~(\ref{eq:key}), as a function of the total length of the link for the three schemes analysed in this section. Here, $n$ is the Maximal ES nesting level. We refer to Tab.~\ref{paraTab} in the Methods section for details about the choice of parameters.}
		\label{fig:long123}
	\end{minipage}

	For this range of distances, maximal ES nesting level $n=2,3$ are optimal, because for the chosen values of the parameters $n\ge4$ gives vanishing key rate. We fix the altitude of the orbits at $h=500$ km in schemes OO and OG. For the latter, at the cost of introducing additional losses, choosing higher orbits has two positive effects: it allows to cover longer distances avoiding the detrimental effect of grazing angle incidence in the atmosphere and makes the fly-by duration longer (see Sec.~\ref{methods2} for details). In scheme OO, instead, going to higher altitudes does not have substantial positive effects.

	The use of orbiting quantum repeater stations clearly gives an important boost to the secret key rate, enlarging at the same time the maximum reachable distance, see Fig.~\ref{fig:long123}. Avoiding the effect of the atmosphere allows to truly take advantage of the quadratic scaling of the losses with the distance that characterizes free-space optical channels in vacuum.
	The proposed scheme OO outperforms schemes GG and OG at every distance beyond $\sim1000$ km, by orders of magnitude. In this case, $n=2$ is enough to achieve non-zero key rate at the longest necessary distance.
	In Fig.~\ref{fig:short} we focus instead on shorter distances, in which scheme OO performs again very well. For the satellite implementations $n=0,1$ are optimal in this case. With $n=0$ schemes OO and OG are identical, as there is just a double down-link to the receiving stations of A and B on the ground \cite{pan4}. In this case, since there are no quantum memories that limit the usable repetition rate, we fix $R_s=1$ GHz. This is the source of the advantage at $L<2000$ km with respect to the other implementations. With $n=1$, scheme OO beats OG by a factor $\sim10$ in this range of distances.
	
	\noindent
	\begin{minipage}{\linewidth}
		\centering
		\includegraphics[width=1\columnwidth]{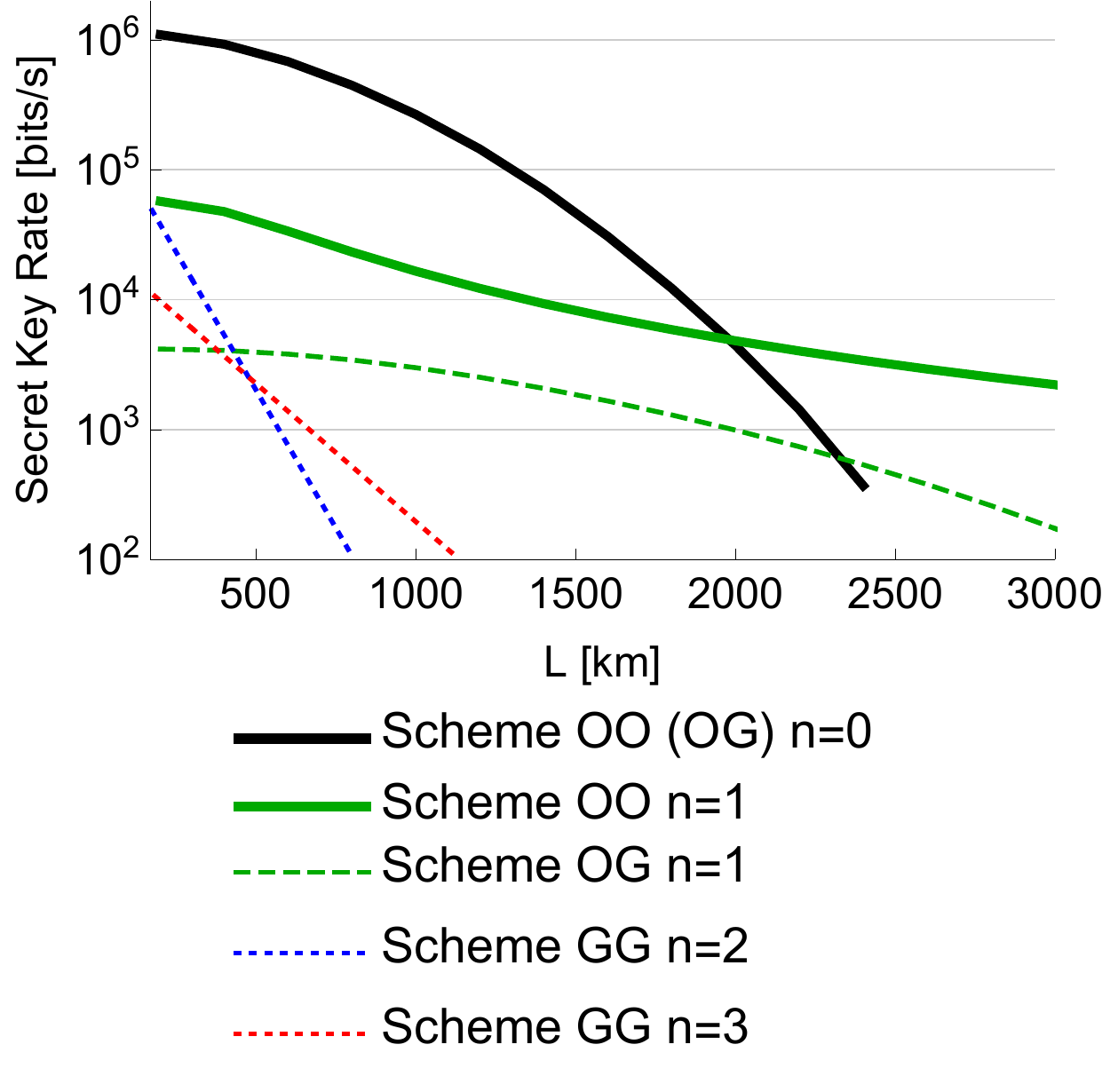}
		\captionof{figure}{The same considerations as in Fig.~\ref{fig:long123} apply, but in this case we focus on short-to-medium distances.}
		\label{fig:short}
	\end{minipage}

	It is important to notice that, while for the ground implementation the link is available all day long, the satellite fly-by duration lasts several minutes at most (see Sec.~\ref{methods2} for details). This means that, if one is interested in the total amount of key exchanged during a day between a specific pair of users A and B, one has to multiply the key rates in Fig.~\ref{fig:long123} and Fig.~\ref{fig:short} by $60 \times 60 \times 24 = 86400$ for the ground implementation and by {50-500} for schemes OO and OG (assuming a single passage). However, as discussed later in Sec.~\ref{examples}, the satellites also give coverage to other regions on Earth, allowing to operate links between different pairs of users in a single orbit (and there are several orbits in one day).
	
	We point out that unlike schemes GG and OG, in scheme OO we find links with different transmittance along the repeater chain, in particular double inter-satellite links and twice an inter-satellite + down-link. The bottleneck given by the link with the lowest transmittance determines the overall entanglement distribution rate. For this reason, some parameters need to be fixed in a smart way. For short distances, the inter-satellite links have high transmittance, so the bottle neck is given by the down-links. In this case, increasing the size of the optics on the repeater satellites is not helpful. For longer distances, instead, the inter-satellite links become longer and lossier, so enlarging the correspondent optics allows to improve the bottleneck.

	\subsection{Pros and cons of orbiting quantum repeater stations}
	\label{results2}
	
	We showed in the previous section how scheme OO of Fig.~\ref{fig:orbiting} reaches the highest key rate at all distances. In this section we will list several additional advantages of this configuration over the other two and discuss the feasibility of its deployment.
	
	First of all, it takes full advantage of inter-satellite-links, which allow to completely avoid the degrading effect of the atmosphere. Even if for down-links the additional diffraction and beam deflections introduced by the atmosphere are generally small \cite{liorni,jennewein1,Bedington}, the inevitable losses due to absorption and backscattering in the air amount to 5-10 dB. In scheme OG, in order for all the links to be active at the same time, good weather conditions must hold in all the intermediate repeater stations. This problem is almost completely solved by scheme OO, for which only the geographical sites of the two parties need to have clear sky conditions. If the channel is divided in $2^n$ elementary links, clear sky conditions must hold in all the $2^n+1$ sites on the ground (A, B and the intermediate repeater stations) for scheme OG. Let us assume that the probability of clear sky in all the locations is $p_{ \text{cs}}$ (uniform and independent). In USA, for example, the sunniest city has $p_{\text{cs}}\simeq0.7$ \cite{ccd2018}, so we assume this value. In this case, for $n=3$, scheme OO gives an additional advantage over scheme OG equal to $p_{\text{cs}}^{-(2^n+1-2)}\simeq 12$. If we analyse Fig.\ref{fig:orbiting}, we see that in scheme OO the satellites need to communicate with a single ground station at a time, unlike scheme OG. For this reason, the fly-by time, that corresponds to the maximum time over which exchange of quantum information is possible, is much longer in scheme OO and independent of the distance between the parties (see Fig.~\ref{fig:flyby} in Sec.~\ref{methods2} for details).
	Finally, while in scheme OO the system is able to link only one pair of parties at a time, the chain of satellites can cover the whole world, depending on the choice of the orbit. In this way, a small number of satellites can guarantee world-wide entanglement distribution, as discussed more thoroughly in Sec.~\ref{examples}.

	The implementation of a full-fledged quantum repeater on a satellite introduces several additional technical challenges with respect to the other schemes. However, in the remainder of this section we will discuss how the main technologies needed have been already developed and tested in the space environment. 
	The low temperature usually needed for the operation of a quantum memory has already been achieved in different experiments. Sub-nK temperatures are expected to be achieved in a trapped atom experiment onboard the International Space Station \cite{cal,cal2}. The same experiment also ensures the ability to reach ultra high vacuum with ease, stable operation of lasers and microwave-radio sources and sizeable artificial magnetic fields. Dilution refrigerators capable of reaching 50 mK with long life-time have already been implemented in micro-gravity conditions \cite{dilution2,dilution3,dilution} and meet the requirements of, for example, quantum memories based on silicon vacancy centres in diamond \cite{lukinmemo,lukinmemo2}.  The first stages of the refrigerator, at $\sim 1$K, can also be shared with Superconducting Nanowire Single-Photon Detectors (SNSPDs).
	The optical inter-satellite links necessary for scheme OO do not comprise important technical difficulties and have already been experimentally realized (e.g., during the SILEX mission of the European Space Agency \cite{silex}).
	In scheme OG the quantum repeater components on the ground could easily be updated over time with newer technology, which is clearly unfeasible in scheme OO. However, we must point out that the life-time of LEO satellites is quite short, few tens of years at most, making it necessary to update the hardware in any case.
	
	\begin{table*}[t]
		\centering
		\begin{tabular}{ | c | c | c | } 
			\hline
			Cities & Distance [km] & Ocean \\ 
			\hline
			New York - San Francisco & 4000 &  \\
			New York - Madrid & 5500 & X \\ 
			San Francisco - Tokyo & 8500 & X \\
			Tokyo - Beijing & 2000 &  \\ 
			Taipei - New Delhi & 4500 &  \\ 
			Hong Kong - Dubai & 6000 & \\
			Beijing - Berlin & 7500 &  \\
			Montreal - Paris & 5500 & X \\
			\hline    
		\end{tabular}
		\medskip
		\caption{\textbf{Exemplary city pairs selection.} Pairs of locations of A and B that we consider as exemplification of the global communication flow, with correspondent distance and whether the link presents long portions in the ocean.}
		\label{cities}
	\end{table*}
	
	\subsection{Analysis of two simple orbital configurations}
	\label{examples}
	
	In this section we qualitatively analyse two types of orbits that may be useful for long-distance entanglement distribution and exemplify the potential of the satellite-based scheme we proposed beforehand. 
	
	The first example consists in Sun-Synchronous orbits, almost polar low Earth orbits that are engineered to pass over a given location always at the same time of the day. These orbits have already been extensively used for all kinds of LEO satellites, from basic research to Earth imaging. Using such orbits, scheme OO would be able to connect cities along the north-south direction. 
	In order to achieve communication in the east-west direction one can use circular orbits with suitable inclination with respect to the equatorial plane, the most promising ones being between $30^\circ$ and $60^\circ$. Such orbits are particularly good to link locations in the temperate and subtropical regions which have roughly the same latitude. If we are interested in serving mostly the northern hemisphere, one can use on-board thrusters to rotate the orbital plane during the year and keep the section of the orbit above the equator in shade. The rotation of the Earth allows every passage of the satellite to cover a different region of the northern hemisphere. In Tab.~\ref{cities} we report an exemplary selection of pairs of important cities, representing the global communication flow, that may be linked by one chain of satellites (using $n=2$ or $n=3$) in a single night. Considering that LEO satellites perform between 16 ($h=500$ km) and 12 ($h=2000$ km) orbits per day, it is reasonable to assume that at least 7-8 long distance links may be achieved. 
	In this case one wants to be able to communicate between different pairs of parties, characterised by different distances, using a configuration of satellites fixed a priori. This means that the number of elementary links $2^n$ and the their length need to be optimized depending on the set of locations we are interested in. One also has some freedom to alter the orbits after the launch and custom the network to the ever-changing needs of the modern society.
	Examining Tab.~\ref{cities}, we see how some of these links would be almost impossible to achieve using scheme OG, that would require optical ground stations in the middle of the ocean, highlighting the advantage of scheme OO.

	\section{Discussion}
	\label{discussion}
	
	In this paper we presented a scheme based on the integration between satellite-based optical links and quantum repeaters to achieve long-distance entanglement distribution and quantum key distribution. 
	Several LEO satellites, carrying quantum sources and quantum repeaters, are linked together by means of inter-satellite optical channels. The end-points of the chain are instead linked to two parties on the ground by down-links. We carefully analyse the repeater rate of the chain and the fidelity of the final shared states, taking into account the effect of different sources of noise. In the end, we compute the asymptotic secret key rate achievable using the BB-84 cryptographic protocol.
	The asymptotic key rate is compared with the rate achievable by an equivalent fibre-based implementation and a different satellite-based configuration \cite{jennewein2}, showing that the proposed scheme significantly outperforms the other approaches and allows the building of a more efficient, reliable and flexible quantum communication network.
	The implementation of the necessary technologies on a satellite has been briefly analysed and seems feasible in the near to mid-term future.
	The ability to share entanglement and distribute keys without relying on trusted nodes differentiate our results from previous works on quantum satellite constellations \cite{network4}. The parameters used in the simulations have been fixed to reasonably conservative values, that should be achievable in the near future.
	Our analysis highlights how for this conservative choice of memory parameters and fidelity the satellite-based configurations with maximal nesting level $n=2$ look more promising than $n=3$ for near- and mid-term implementation. For better memories and sources the additional round of entanglement swapping would be less costly and the reduced losses in the elementary pairs would allow for higher rates.
	A key factor in deciding the best solution for entanglement distribution on a global scale is the cost, that we qualitatively estimate in the following. In order to compare the different schemes we assume the pairs of cities in Tab.~\ref{cities} as a benchmark. For scheme GG, we assume that all the pairs of cities are connected by separate fibre links. We take as base price the cost of the Shanghai-Beijing quantum network ($\sim 100$ million euro for 2000 km total length, based on trusted repeaters, \cite{chinabackbone}). This gives a total of 2.2 billion euro, without considering the additional costs related to the quantum repeaters and the fact that links under the ocean would be more expensive to install. For scheme OG, considering $n=2$, we need 4 source satellites, 3 intermediate repeater stations for every link 
	and one receiving station per city (12). Assuming 50 million euro per satellite \cite{satcost} (Micius costed $\sim 100$ million dollar including development \cite{miciuscost}), 2 million per ground station \cite{OGScost} and 20 million per ground station in the ocean, we reach a total of 430 million euro.  
	For scheme OO, considering again $n=2$, we need 4 source satellites, 3 repeater satellites and one receiving station per city (12). We assume again 50 million euro per source satellite, 150 million per repeater satellite \cite{satcost} and 2 million per ground station, summing up to 674 million euro. 
	We deduce that scheme GG is the most expensive one and also gives completely unsatisfying key rate, calling for the use of different quantum repeater architectures \cite{rep3}. Scheme OO is about $50\%$ more expensive than scheme OG, but gives a total advantage in performance of a factor $\sim1000$, considering both higher key rate and availability of the link.
	In summary, the global quantum channels analysed in this work, built through the integration of satellite-based links and repeater nodes, can be envisaged to represent the backbone of the future Quantum Internet \cite{network2,network1,network3,qinternet}.

	\section{Methods} 
	\label{methods}

	\subsection{Error model and environmental photons}
	\label{methods1}
	
	In this section we will discuss additional aspects regarding the noise model used for the simulations of Sec.~\ref{results}.
	In order to compute the probabilities $P_{s1}$ and $P_{s2}$ of Eq.~(\ref{FFF}) we need an estimate of the number of environmental photons per time window at the receiver. In the case of scheme OG all the receivers are on the ground and we can consider the same background light for every site. We assume that the receiving telescope has radius $r$, field of view $\Omega_{\text{fov}}$ and that we apply spectral and temporal filtering with widths $B_f$ and $\Delta t=1/R_s$.
	If the artificial light pollution is negligible, the power received by the telescope can be expressed as follows \cite{miao}
	\begin{equation}
	\label{p}
	P_{\text{noise}} = H_b \Omega _{\rm fov} \pi r^2 B_f \ .
	\end{equation}
	The parameter $H_b$ is the total brightness of the sky background and it depends on the hour of the day and the weather conditions. From Eq.~(\ref{p}) we derive the number of photons per time window 
	\begin{equation}
	\label{noise}
	N_{\text{noise}}=\frac{H_b}{h \nu} \Omega_{\rm fov} \pi a^2 B_f \ \Delta t \ ,
	\end{equation}
	where $h$ is the Planck constant and $\nu$ is the frequency of the background photons. Typical values of the brightness of the sky at the wavelength under study are $H_b=10^{-3} \ W \ {\rm m}^{-2} \ {\rm sr} \ \mu {\rm m}$ during a full-Moon night (this value has been used in the simulations in the main text) and $H_b=1 \ W \ {\rm m}^{-2} \ {\rm sr} \ \mu {\rm m}$ for a clear sky in day-time. 
	
	In scheme OO we have receivers on the ground (at the parties A and B) and in LEO. The latter ones are used in inter-satellite links, so they are pointing towards the adjacent satellites, in a direction more or less tangent to the Earth's surface and atmosphere. Due to the narrow field of view, they will receive practically no light reflected or diffused from the Earth and the atmosphere. The background light from celestial objects should be negligible and so should be any reflection coming from the sending satellite \cite{bonato}. This means that the intermediate repeater nodes will be affected by almost no additional noise and only the photons that are sent towards parties A and B at the two ends of the chain will mix with environmental light. However, in order to simplify the analysis, we assume in the simulations that all the photon pairs have the same noise level as the ones comprising the down-link, getting a lower bound on the final secret key rate.
	
	Another source of errors is represented by dark counts in the detectors used for the BSM. We assume here the standard linear optics setup for polarization-entanglement, in which the photons read from the memories are let interfere on a beam splitter. The light coming out of the two output ports is then analysed using two polarizing beam splitters and 4 single photon detectors. The different click patterns allow to distinguish two out of the four possible Bell states in input. In this case the success probability of the entanglement swapping procedure $P_{ES}$, used in Eq.~(\ref{eq:rep}) of the main text, can be expressed as \cite{abruzzo}
	\begin{equation}
	\label{pes}
	P_{ES}=\frac{1}{2}\{ [1-p_{\text{dark}}][\eta_d+2 \ p_{\text{dark}}(1-\eta_d)] \}^2 \ ,
	\end{equation}
	where $p_{\text{dark}}$ is the detector dark count probability and $\eta_d$ their efficiency.
	
	So far we considered the imperfections of the quantum memories limited to non-unity writing and reading efficiencies. Decoherence in the memories should be addressed too, especially because very long distances beyond 10000 km correspond to long communication times of tens of ms. As discussed in \cite{jennewein2}, such long coherence times should be achievable by transferring the optical memory excitations to the ground spin states, for example in systems based on Eu-doped yttrium orthosilicate. Electronic spin states can be transferred to long-lived nuclear spin states in silicon-vacancy centres in diamond. In our simulation, such a modification would correspond to a lower value of the writing efficiency $P_W$ and would act in the same way on the different implementations, not changing the comparison between them.
	
	\subsection{Modelling the orbits and the transmittance of the satellite links}
	\label{methods2}

	In this section we will give some details about the orbit model and how the transmittance of the satellite-based optical links has been computed. We assume circular orbits at altitude $h$ above the ground and that, for simplicity, they lie in the equatorial plane. The ground stations are likewise put along the equator. The results of the paper can be extended to repeater chains in different sites of the globe by using suitable orbits (e.g., Sun Synchronous LEO). The law of motion of the satellites and the relative position with respect to the ground stations have then been computed using simple geometrical considerations and the law of gravitational force, without any relativistic correction. In scheme OG, we define the fly-by as the period of time during which the satellite is in line-of-sight contact with both the adjacent ground stations. To be in contact, we suppose that it must be at an elevation angle, in the local coordinate frame of the ground stations, greater than a threshold that we set to $15^\circ$ \cite{Liao}. The duration of the fly-by depends on the altitude of the satellite (that also fixes the angular speed), on the orbital direction (the same or opposite to the rotation of the Earth) and on the distance between the ground stations, fixed by the total distance $L$ and $n$. 
	
	\noindent 
	\begin{minipage}{\linewidth}
		\centering		\includegraphics[width=1\columnwidth]{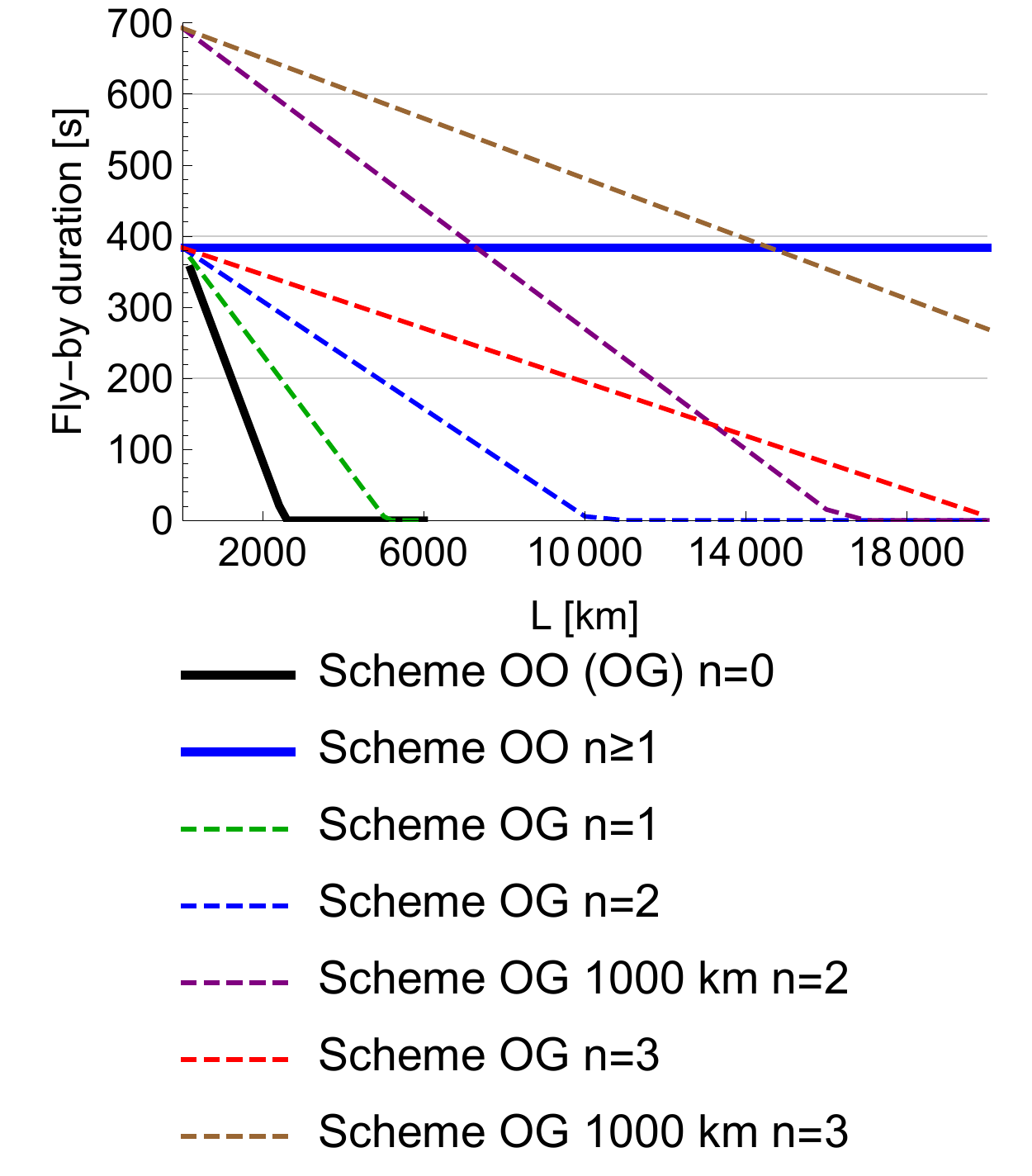}
		\captionof{figure}{Fly-by duration as a function of the total distance between A and B, for different values of the maximal ES nesting level $n$ and altitude (fixed at 500 km where not specified). Notice that for scheme OO the duration is independent of $L$ and $n\geq1$.}
		\label{fig:flyby}
	\end{minipage}

	The effect is shown in Fig.~\ref{fig:flyby}, where it is evident how the fly-by duration goes to 0 when the distance between the ground stations becomes too large. This is not true for scheme OO, where it only depends on the altitude and it is independent of the distance $L$ for $n\geq1$.

	In the remainder of this section we will outline the methodology used to estimate the instantaneous value of the transmittance of the free-space links. The beam effects introduced by the atmosphere \cite{liorni,jennewein1,Bedington}, like additional beam wandering and broadening, are neglected in this work, as their effect is small compared to the strong geometrical losses due to the intrinsic diffraction. Same holds for losses related to pointing inaccuracy. We assume that the transmitter on the satellite generates a collimated imperfect Gaussian beam with initial beam waist $W_0$ and quality factor $M^2$ \cite{gaussianm2}. The value of the parameter $M^2$ has been fixed to match the far-field divergence of the imperfect Gaussian beam to the one observed for the mission Micius \cite{Liao,pan2,pan3,pan4}. If we suppose that smaller values of $M^2$ can be achieved (better correction of optical aberrations) the value of the transmittance of the free-space links can easily go up of a factor $\{5-10\}$. 
	
	The atmosphere introduces losses due to absorption and back-scattering that depend on the elevation angle $\theta$ of the source and the frequency of the light. 
	We fix the wavelength $\lambda=580$ nm, the operating wavelength of Eu-doped yttrium orthosilicate memories \cite{multimode}, also a good compromise considering atmospheric extinction and diffraction.
	
	The beam waist of a collimated imperfect Gaussian beam will broaden during the propagation in vacuum, following the relation \cite{salehteich}
	\begin{equation}
	\label{wz}
	W(z)=W_0\sqrt{1+(zM^2/z_R)^2} \ .
	\end{equation}
	In the far field limit $z\gg z_R/M^2$, with $z_R=\pi W_0^2/\lambda$ the Rayleigh parameter of the beam with wavelength $\lambda$, Eq.~(\ref{wz}) is linear in the distance $z$. Now we compute the integral of the Gaussian intensity distribution at the receiver, with beam waist $W(z=\bar{z})$, inside a circular region with radius $R$, obtaining
	\begin{equation}
	\label{etadiffr}
	\eta_{\text{diffr}}(\bar{z})=1-\text{exp}\bigg[ -2 \frac{R}{W^2(\bar{z})} \bigg] \ .
	\end{equation}
	This corresponds with the transmittance of the imperfect Gaussian beam through the receiving aperture of radius $R$, when the beam is perfectly aligned and centred. This formula can be directly employed for the inter-satellite links of scheme OO, while we multiply it by the factor $\chi_{\text{ext}}(\theta)=\text{exp}[-\beta \sec(\theta)]$ to take into account atmospheric extinction. $\beta$ depends on the site and the atmospheric condition (see \cite{jennewein1} for details).

	\begin{table*}[t]
		\centering
		\begin{tabular}{ | c | c | c | c| c|} 
			\hline
			Parameter & Value & Brief description & Eq./Sec. & Ref.\\ 
			\hline
			$P_{\text{dark}}$ & $10 ^{-5}$ & Detector dark click probability & Eq.~(\ref{pes}) & \cite{eff1,eff2}\\  
			\hline
			$\eta_d$ & $0.9$ & Detector efficiency & Eq.~(\ref{eq:P}) & \cite{eff1,eff2}\\
			\hline
			$P_W$ & 0.9 & Memory writing efficiency & Eq.~(\ref{eq:rep}) & \cite{abruzzo,jennewein2,memoeff}\\
			\hline 
			$P_R$ & 0.9 & Memory reading efficiency & Eq.~(\ref{eq:rep}) & \cite{abruzzo,jennewein2,memoeff}\\ 
			\hline
			$P_{QND}$ & 0.5 & QND measurement efficiency & Eq.~(\ref{eq:rep}) & \cite{abruzzo,jennewein2,memoeff}\\
			\hline
			$R_s$ & 20 MHz & Repetition rate of the source & Sec.~\ref{results1} & \cite{jennewein2,memoeff}\\
			\hline
			$R_s^{\text{direct}}$ & 1 GHz & Repetition rate for direct transmission & Sec.~\ref{results1} & \cite{jennewein2,memoeff}\\
			\hline
			$\alpha$ & 0.17 dB/km & Fibre loss coefficient at 1550nm & Sec.~\ref{results1} & \cite{abruzzo}\\
			\hline
			$W_0$ & 0.25 m & Beam width at the transmitter & Eq.~(\ref{wz}) & \cite{Liao,jennewein1,jennewein2}\\
			\hline
			$R_{OO}$ & 0.5 m & Radius of the receiver telescope, scheme OO & Eq.~(\ref{etadiffr}) & \cite{Liao,jennewein1,jennewein2}\\
			\hline
			$R_{OG}$ & 0.5 m & Radius of the receiver telescope, scheme OG & Eq.~(\ref{etadiffr}) & \cite{Liao,jennewein1,jennewein2}\\
			\hline
			$\lambda$ & 580 nm & Wavelength of the light, schemes OO and OG & Sec.~\ref{methods2} & \cite{jennewein2,multimode}\\
			\hline
			$M^2$ & 3 & Quality factor of the Gaussian beams & Eq.~(\ref{wz}) & \cite{Liao,gaussianm2}\\
			\hline
			$\beta$ & 1.1 & Atmospheric extinction parameter at 580nm & Sec.~\ref{methods2} & \cite{jennewein1}\\ 
			\hline
			$F_0$ & 0.98 & Initial pair fidelity & Eq.~(\ref{FFF}) & \cite{abruzzo}\\ 
			\hline
			$H_b$ & $1.5$ $\mu W\text{m}^{-2} \ \text{sr}^{-1} \text{nm}^{-1}$ & Total brightness of the sky background & Eq.~(\ref{noise}) & \cite{miao,qdots}\\ 
			\hline
			$\Omega_{\text{fov}}$ & $(20 \ 10^{-6})^2 \text{sr}$ & Field of view of the receiver & Eq.~(\ref{noise}) & \cite{miao,qdots}\\ 
			\hline
			$B_f$ & 0.5 nm & Spectral filter bandwidth & Eq.~(\ref{noise}) & \cite{miao,qdots}\\ 
			\hline
			$\Delta T$ & $1/R_s$ & Time filter bandwidth & Eq.~(\ref{noise}) & \cite{miao,qdots}\\ 
			\hline
		\end{tabular}
		\medskip
		\caption{Parameters used in all the simulations in Sec.~\ref{results}. The parameters have been chosen to represent a reasonable prediction of what can be achieved in the near future. Superconducting Nanowire Single-Photon Detectors (SNSPDs) with low dark count rates and efficiencies exceeding $90\%$ have already been realized at different wavelengths \cite{eff1,eff2}. The quantum memory and heralding parameters have been already used in other theoretical studies \cite{abruzzo,jennewein2} and the recent developments in the field make them reasonable \cite{memoeff}. The size of the optical elements is a plausible improvement over past experiments \cite{Liao,pan2,pan3,pan4} and is also compatible with low-cost optical ground stations. The parameters regarding the environmental light filtering should be reasonably easy to achieve and even improve \cite{miao,qdots}.}
		\label{paraTab}
	\end{table*}
	
	The instantaneous value of the transmittance of the double link from the source to the adjacent repeater stations is then averaged over the fly-by and this quantity is used in Eq.~(\ref{eq:rep}) of the main text, labelled as $P_0$. Scheme OO contains two types of links, double inter-satellite links and twice an inter-satellite + down-link. For every configuration we compare the transmittance of the two types of links and used as $P_0$ the smaller one, that represents the bottleneck in the chain. 
	
	In Tab.~\ref{paraTab} we report the values of the most important parameters used in the simulations of Sec.~\ref{results}.

	\bibliographystyle{unsrt}
	\bibliography{satelliteComparison}

\begin{thebibliography}{10}

\bibitem{gisinreview}
N.~Gisin and R.~Thew.
\newblock Quantum communication.
\newblock {\em Nature Photonics}, 1(3):165--171, 2007.

\bibitem{krenn}
M.~Krenn, M.~Malik, T.~Scheidl, R.~Ursin, and A.~Zeilinger.
\newblock {\em Quantum Communication with Photons}, pages 455--482.
\newblock Springer International Publishing, Cham, 2016.

\bibitem{metrology1}
V.~Giovannetti, S.~Lloyd, and L.~Maccone.
\newblock Advances in quantum metrology.
\newblock {\em Nature Photonics}, 5(4):222--229, 2011.

\bibitem{metrology2}
G.~T{\'{o}}th and I.~Apellaniz.
\newblock Quantum metrology from a quantum information science perspective.
\newblock {\em Journal of Physics A: Mathematical and Theoretical},
  47(42):424006, oct 2014.

\bibitem{metrology3}
E.~T. Khabiboulline, J.~Borregaard, K.~De~Greve, and M.~D. Lukin.
\newblock Optical interferometry with quantum networks.
\newblock {\em Phys. Rev. Lett.}, 123:070504, Aug 2019.

\bibitem{distr1}
R.~{Van Meter} and S.~J. {Devitt}.
\newblock The path to scalable distributed quantum computing.
\newblock {\em Computer}, 49(9):31--42, Sep. 2016.

\bibitem{distr2}
A.~Yimsiriwattana and S.~J.~Lomonaco Jr.
\newblock {Distributed quantum computing: a distributed Shor algorithm}.
\newblock In Eric Donkor, Andrew~R. Pirich, and Howard~E. Brandt, editors, {\em
  Quantum Information and Computation II}, volume 5436, pages 360 -- 372.
  International Society for Optics and Photonics, SPIE, 2004.

\bibitem{repeater1}
H.~J. Briegel, J.~I. Cirac, W.~D{\"u}r, G.~Giedke, and P.~Zoller.
\newblock {\em Quantum Repeaters for Quantum Communication}, pages 147--154.
\newblock Springer Netherlands, Dordrecht, 1999.

\bibitem{rep5}
N.~Sangouard, C.~Simon, H.~de~Riedmatten, and N.~Gisin.
\newblock Quantum repeaters based on atomic ensembles and linear optics.
\newblock {\em Rev. Mod. Phys.}, 83:33--80, Mar 2011.

\bibitem{rep1}
W.~J. {Munro}, K.~{Azuma}, K.~{Tamaki}, and K.~{Nemoto}.
\newblock Inside quantum repeaters.
\newblock {\em IEEE Journal of Selected Topics in Quantum Electronics},
  21(3):78--90, May 2015.

\bibitem{rep2}
K.~Azuma, K.~Tamaki, and H.-K. Lo.
\newblock All-photonic quantum repeaters.
\newblock {\em Nature Communications}, 6:6787 EP --, Apr 2015.

\bibitem{rep3}
M.~Zwerger, A.~Pirker, V.~Dunjko, H.~J. Briegel, and W.~D\"ur.
\newblock Long-range big quantum-data transmission.
\newblock {\em Phys. Rev. Lett.}, 120:030503, Jan 2018.

\bibitem{rep4}
Z.~Su, J.~Guan, and L.~Li.
\newblock Efficient quantum repeater with respect to both
  entanglement-concentration rate and complexity of local operations and
  classical communication.
\newblock {\em Phys. Rev. A}, 97:012325, Jan 2018.

\bibitem{qmem1}
A.~I. Lvovsky, B.~C. Sanders, and W.~Tittel.
\newblock Optical quantum memory.
\newblock {\em Nature Photonics}, 3(12):706--714, 2009.

\bibitem{ecrep}
S.~Muralidharan, J.~Kim, N.~L\"utkenhaus, M.~D. Lukin, and L.~Jiang.
\newblock Ultrafast and fault-tolerant quantum communication across long
  distances.
\newblock {\em Phys. Rev. Lett.}, 112:250501, Jun 2014.

\bibitem{Liao}
S.-K. Liao et~al.
\newblock Satellite-to-ground quantum key distribution.
\newblock {\em Nature}, 549:43 EP --, Aug 2017.
\newblock Article.

\bibitem{pan2}
J.~Yin et~al.
\newblock Satellite-to-ground entanglement-based quantum key distribution.
\newblock {\em Phys. Rev. Lett.}, 119:200501, Nov 2017.

\bibitem{pan3}
J.-G. Ren et~al.
\newblock Ground-to-satellite quantum teleportation.
\newblock {\em Nature}, 549:70 EP --, Aug 2017.

\bibitem{pan4}
J.~Yin et~al.
\newblock Satellite-based entanglement distribution over 1200 kilometers.
\newblock {\em Science}, 356(6343):1140--1144, 2017.

\bibitem{bonato}
C.~Bonato, A.~Tomaello, V.~Da Deppo, G.~Naletto, and P.~Villoresi.
\newblock Feasibility of satellite quantum key distribution.
\newblock {\em New Journal of Physics}, 11(4):045017, 2009.

\bibitem{jennewein1}
J.-P. Bourgoin et~al.
\newblock A comprehensive design and performance analysis of low earth orbit
  satellite quantum communication.
\newblock {\em New Journal of Physics}, 15(2):023006, 2013.

\bibitem{liorni}
C.~Liorni, H.~Kampermann, and D.~Bru{\ss}.
\newblock Satellite-based links for quantum key distribution: beam effects and
  weather dependence.
\newblock {\em New Journal of Physics}, 21(9):093055, sep 2019.

\bibitem{abruzzo}
S.~Abruzzo et~al.
\newblock Quantum repeaters and quantum key distribution: Analysis of
  secret-key rates.
\newblock {\em Phys. Rev. A}, 87:052315, May 2013.

\bibitem{jennewein2}
K.~Boone et~al.
\newblock Entanglement over global distances via quantum repeaters with
  satellite links.
\newblock {\em Phys. Rev. A}, 91:052325, May 2015.

\bibitem{rep0}
H.-J. Briegel, W.~D\"ur, J.~I. Cirac, and P.~Zoller.
\newblock Quantum repeaters: The role of imperfect local operations in quantum
  communication.
\newblock {\em Phys. Rev. Lett.}, 81:5932--5935, Dec 1998.

\bibitem{bb84}
C.H. Bennett and G.~Brassard.
\newblock Quantum cryptography: Public key distribution and coin tossing.
\newblock {\em Proceedings of IEEE International Conference on Computers,
  Systems, and Signal Processing, Bangalore}, pages 175--179, 1984.

\bibitem{deRiedmatten}
H.~de~Riedmatten, M.~Afzelius, M.~U. Staudt, C.~Simon, and N.~Gisin.
\newblock A solid-state light-matter interface at the single-photon level.
\newblock {\em Nature}, 456(7223):773--777, 2008.

\bibitem{multimode}
M.~Afzelius, C.~Simon, H.~de~Riedmatten, and N.~Gisin.
\newblock Multimode quantum memory based on atomic frequency combs.
\newblock {\em Phys. Rev. A}, 79:052329, May 2009.

\bibitem{lukinmemo}
M.~K. Bhaskar et~al.
\newblock Experimental demonstration of memory-enhanced quantum communication,
  2019.

\bibitem{vanlock}
N.~K. Bernardes, L.~Praxmeyer, and P.~van Loock.
\newblock Rate analysis for a hybrid quantum repeater.
\newblock {\em Phys. Rev. A}, 83:012323, Jan 2011.

\bibitem{renner1}
R.~Renner, N.~Gisin, and B.~Kraus.
\newblock Information-theoretic security proof for quantum-key-distribution
  protocols.
\newblock {\em Phys. Rev. A}, 72:012332, Jul 2005.

\bibitem{bennett1}
C.~H. Bennett, D.~P. DiVincenzo, J.~A. Smolin, and W.~K. Wootters.
\newblock Mixed-state entanglement and quantum error correction.
\newblock {\em Phys. Rev. A}, 54:3824--3851, Nov 1996.

\bibitem{distillation}
C.~H. Bennett, H.~J. Bernstein, S.~Popescu, and B.~Schumacher.
\newblock Concentrating partial entanglement by local operations.
\newblock {\em Phys. Rev. A}, 53:2046--2052, Apr 1996.

\bibitem{Bedington}
R.~Bedington, J.~M. Arrazola, and A.~Ling.
\newblock Progress in satellite quantum key distribution.
\newblock {\em npj Quantum Information}, 3(1):30, 2017.

\bibitem{ccd2018}
Comparative climatic data for the united states through 2018.
\newblock https://www.ncdc.noaa.gov/ghcn/comparative-climatic-data.
\newblock Accessed: 2020-03-04.

\bibitem{cal}
E.~R. Elliott et~al.
\newblock Nasa's cold atom lab (cal): system development and ground test
  status.
\newblock {\em npj Microgravity}, 4(1):16, 2018.

\bibitem{cal2}
T.~Schuldt et~al.
\newblock Design of a dual species atom interferometer for space.
\newblock {\em Experimental Astronomy}, 39(2):167--206, Jun 2015.

\bibitem{dilution2}
S.~Triqueneaux, L.~Sentis, Ph. Camus, A.~Benoit, and G.~Guyot.
\newblock Design and performance of the dilution cooler system for the planck
  mission.
\newblock {\em Cryogenics}, 46(4):288 -- 297, 2006.

\bibitem{dilution3}
G.~Chaudhry, A.~Volpe, P.~Camus, S.~Triqueneaux, and G.~Vermeulen.
\newblock A closed-cycle dilution refrigerator for space applications.
\newblock {\em Cryogenics}, 52:471–477, 10 2012.

\bibitem{dilution}
M.~Zheng et~al.
\newblock A brief review of dilution refrigerator development for space
  applications.
\newblock {\em Journal of Low Temperature Physics}, 197(1):1--9, Oct 2019.

\bibitem{lukinmemo2}
C.~T. Nguyen et~al.
\newblock Quantum network nodes based on diamond qubits with an efficient
  nanophotonic interface.
\newblock {\em Phys. Rev. Lett.}, 123:183602, Oct 2019.

\bibitem{silex}
G.~D. {Fletcher}, T.~R. {Hicks}, and B.~{Laurent}.
\newblock The silex optical interorbit link experiment.
\newblock {\em Electronics Communication Engineering Journal}, 3(6):273--279,
  Dec 1991.

\bibitem{network4}
T.~Vergoossen et~al.
\newblock {Satellite constellations for trusted node QKD networks}.
\newblock {\em arXiv e-prints}, page arXiv:1903.07845, Mar 2019.

\bibitem{chinabackbone}
Beijing-shanghai quantum communication network put into use.
\newblock
  \url{http://english.cas.cn/newsroom/archive/news_archive/nu2017/201703/t20170324_175288.shtml}.
\newblock Accessed: 2020-03-04.

\bibitem{satcost}
H.~Stahl, F.~Prince, C.~Smart, K.~Stephens, and T.~Henrichs.
\newblock Preliminary cost model for space telescopes.
\newblock {\em Proc SPIE}, 7436, 08 2009.

\bibitem{miciuscost}
K.E. Lee and A.~Khor.
\newblock {\em China’s Long March of Modernisation: Blueprint \& Road Map for
  the Nation’s Full Development 2016-2049}.
\newblock Xlibris AU, 2019.

\bibitem{OGScost}
E.~Clements, J.~L. Mendenhall, D.~Caplan, and K.~Cahoy.
\newblock Lasercom uncertainty modeling and optimization simulation (lumos): A
  statistical approach to risk-tolerant systems engineering for small
  satellites.
\newblock 2019.

\bibitem{network2}
S.~Wehner, D.~Elkouss, and R.~Hanson.
\newblock Quantum internet: A vision for the road ahead.
\newblock {\em Science}, 362(6412), 2018.

\bibitem{network1}
S.~Wengerowsky et~al.
\newblock An entanglement-based wavelength-multiplexed quantum communication
  network.
\newblock {\em Nature}, 564(7735):225--228, 2018.

\bibitem{network3}
A.~Dahlberg et~al.
\newblock {A Link Layer Protocol for Quantum Networks}.
\newblock {\em arXiv e-prints}, Mar 2019.

\bibitem{qinternet}
M.~Pant, H.~Krovi, D.~Towsley, L.~Tassiulas, L.~Jiang, P.~Basu, D.~Englund, and
  S.~Guha.
\newblock Routing entanglement in the quantum internet.
\newblock {\em npj Quantum Information}, 5(1):25, 2019.

\bibitem{miao}
E.-L. Miao, Z.-F. Han, S.-S. Gong, T.~Zhang, D.-S. Diao, and G.-C. Guo.
\newblock Background noise of satellite-to-ground quantum key distribution.
\newblock {\em New Journal of Physics}, 7(1):215, 2005.

\bibitem{gaussianm2}
A.~E. Siegman.
\newblock {Defining, measuring, and optimizing laser beam quality}.
\newblock In Anup Bhowmik, editor, {\em Laser Resonators and Coherent Optics:
  Modeling, Technology, and Applications}, volume 1868, pages 2 -- 12.
  International Society for Optics and Photonics, SPIE, 1993.

\bibitem{salehteich}
B.E.A. Saleh and M.C. Teich.
\newblock {\em Fundamentals of photonics}.
\newblock Wiley series in pure and applied optics. Wiley, 1991.

\bibitem{eff1}
A.~E. Lita, A.~J. Miller, and S.~W. Nam.
\newblock Counting near-infrared single-photons with 95\% efficiency.
\newblock {\em Opt. Express}, 16(5):3032--3040, Mar 2008.

\bibitem{eff2}
D.~Fukuda et~al.
\newblock Titanium-based transition-edge photon number resolving detector with
  98\% detection efficiency with index-matched small-gap fiber coupling.
\newblock {\em Opt. Express}, 19(2):870--875, Jan 2011.

\bibitem{memoeff}
J.~Guo.
\newblock High-performance raman quantum memory with optimal control in room
  temperature atoms.
\newblock {\em Nature Communications}, 10(1):148, 2019.

\bibitem{qdots}
T.~T. Tran et~al.
\newblock Nanodiamonds with photostable, sub-gigahertz linewidth quantum
  emitters.
\newblock {\em APL Photonics}, 2(11):116103, 2017.

\end{thebibliography}
	
	\section{Acknowledgements}
	
	This project has received funding from the European Union's Horizon 2020 research and innovation programme under the Marie Sk\l{}odowska-Curie grant agreement No. 675662. D.B. and H.K. acknowledge support from the Federal Ministry of Education and Research BMBF (Project Q.Link.X).

	\section{Author contributions}
	
	C. L. conceived the work and performed the simulations. D. B. and H. K. reviewed the analysis. All the authors discussed the results and contributed to the final manuscript. 
	
	\section{Additional Information}
	
	\textbf{Competing financial interests:} The authors declare no competing financial interests.
	
	\end{multicols}

\end{document}